\documentclass[12pt]{report}
\usepackage[dvips]{graphicx}
\newcommand{\sptwo}{1.4}

\newcommand{\doublespace}{\edef\baselinestretch{\sptwo}\Large\normalsize}

\begin{document}
\doublespace
\begin{center}
{\bf  Hydrodynamic Modes in a Trapped  Strongly Interacting Fermi Gases of Atoms }\\
\renewcommand\thefootnote{\fnsymbol{footnote}}
{Yeong E. Kim \footnote{ e-mail:yekim$@$physics.purdue.edu} and
Alexander L. Zubarev\footnote{ e-mail: zubareva$@$physics.purdue.edu}}\\
Purdue Nuclear and Many-Body Theory Group (PNMBTG)\\
Department of Physics, Purdue University\\
West Lafayette, Indiana  47907\\
\end{center}

\begin{quote}
The zero-temperature properties of a dilute two-component Fermi gas in the
BCS-BEC crossover are investigated. On the basis of a generalization of the
variational Schwinger method,
 we construct approximate  semi-analytical formulae for
  collective
 frequencies of the radial and the axial breathing modes of the Fermi gas
 under harmonic confinement in the framework of the
 hydrodynamic theory. 
It is shown that the method gives nearly exact solutions.
%the absolute precision of the method is expected to be of order $10^{-5}$.

%while the precision  of the scaling approximation
%  is about 0.5 \%.

\end{quote}

\vspace{5mm}
\pagebreak

\noindent
{\bf 1. Introduction}
\vspace{8pt}

The newly created ultracold trapped Fermi gases with tunable atomic scattering
 length [1-20] in the vicinity of  a Feshbach resonance offer the possibility to
 study highly
 correlated many-body systems including the cross-over from the
 Bardeen-Cooper-Schrieffer (BCS) phase to the Bose-Einstein condensate (BEC)
 of molecules.  Various investigations  based on the hydrodynamic theory have
appeared recently [21-31].

The purpose of this letter is to construct a simple, semi-analytical and
nearly  exact formulae for hydrodynamic frequencies. Since 
 the collective frequencies can be measured with high precision,
 these formulae will provide
a simple quantitative tool for the analysis of experimental data in
the hydrodynamic regime.

\vspace{8pt}

\noindent
{\bf 2. Hydrodynamic theory}
\vspace{8pt}

Our starting point is the quantum hydrodynamic theory [21-23] 
for a dilute two-component Fermi gas in a
trap potential $V_{ext}(\vec{r})=(m/2)(\omega_{\perp}^2 (x^2+y^2)+
\omega_z^2 z^2)$
 $$
i\hbar \frac{\partial \Psi}{\partial t}=-\frac{ \hbar^2}{2 m} \nabla^2 \Psi
+V_{ext} \Psi+V_{xc}\Psi,
\eqno{(1)}
$$
where
$$V_{xc}(\vec{r},t)=[\frac{\partial( n \epsilon(
n))}{\partial n}]_{n=n(\vec{r},t)},$$ $\epsilon$
is the ground state energy per particle of the homogeneous system and $n$ is the
 density, $n(\vec{r},t)=\mid \Psi(\vec{r},t) \mid^2$, 
normalized to the total number of atoms, $\int n(\vec{r},t) d^3r=N$.
It is useful to rewrite Eq.(1) in a  form
$$
\frac{\partial n}{\partial t}+\nabla (n \vec{\mbox{v}})=0,
\eqno{(2)}
$$
$$
\frac{\partial \vec{\mbox{v}}}{\partial t}+\frac{1}{m}\nabla (V_{ext}+
\frac{d(n \epsilon(
n))}{d n}+\frac{1}{2} m \mbox{v}^2-\frac{\hbar^2}{2 m} \frac{1}{\sqrt{n}}
 \nabla^2 \sqrt{n})=0,\eqno{(3)}
$$
where $\vec{\mbox{v}}$ is the velocity field, which for 
$\Psi=e^{i \phi(\vec{r},t)} n^{1/2}(\vec{r},t)$ can be written as
 $\vec{\mbox{v}}=(\hbar/m) \nabla \phi$.

It can be proved [21] that every solution of the Eqs.(2,3) is a stationary point corresponding to the Lagrangian density
$$
\mathcal{L}_0=\hbar \dot{\phi}n+\frac{\hbar^2}{2m} (\nabla \sqrt{n})^2+
\frac{\hbar^2}{2m}n (\nabla \phi)^2+\epsilon(n)n+V_{ext}n.
\eqno{(4)}
$$
It was shown in Refs.[21-23] that for experimental conditions of
 Refs.[14,16,17]
the quantum pressure term in Eqs.(3,4)  can be neglected. For the reminder of
 this
Letter we
 will use this hydrodynamic approximation.
For the harmonic trap a trial function in the scaling ansatz is taken as
 [21-23,26,28]
$$\phi(\vec{r},t)=\phi_0(t)+(m/(2 \hbar) \sum_{i=1}^3 \beta_i(t) x_i^2,$$
$$n(\vec{r},t)=n_0(x_i/b_i(t))/\kappa(t),$$ where $\kappa(t)=\prod_jb_j$ and
the Hamilton principle, $\delta\int dt\int\mathcal{L}_0 d^3 r=
0$, gives the following equations for the scaling parameters $b_i$ [21-22]
$$
\ddot{b}_i+\omega_i^2(t) b_i-\frac{\omega_i^2}{b_i} \frac
{\int[n^2d\epsilon(n)/dn]_{n=n_0(\vec{r})/\kappa(t)} d^3 r}
{\int[n^2d\epsilon(n)/dn]_{n=n_0(\vec{r})} d^3 r}\kappa(t)=0,
\eqno{(5)}
$$

 Expanding Eqs.(5) around equilibrium ($b_i=1$)   leads  
to
the following result for the $M=0$ modes frequencies, $\omega^{(s)}$ 
in the scaling approximation
$$
\omega^{(s)}_{\pm}=\frac{\omega_{\perp}}{\sqrt{2}} 
[\eta_s \pm \sqrt{\eta_s^2-8 \lambda^2 (3 \zeta_s+5)}]^{1/2},
 \eqno{(6)}
$$
where $\eta_s=4+2\zeta_s+3 \lambda^2+\zeta_s \lambda^2$, $\zeta_s=
\int n_0^3d^2\epsilon/(dn_0^2)d^3r/\int n_0^2d\epsilon/(d  n_0)d^3r$,
$\lambda=\omega_z/\omega_{\perp}$, and $\pm$ refer to the transverse and axial
mode, respectively.

The hydrodynamic equations after linearization take the form
$$
\frac{\partial^2}{\partial t^2} \delta n+\frac{1}{m} \nabla (
n_0 \nabla (\frac{d^2(n_0 \epsilon(n_0)))}{d n_0^2} \delta n))=0,
\eqno{(7)}
$$
where $\delta n(\vec{r},t)$ is the change in the density profile with respect
to the equilibrium configuration. If we consider oscillations with time
 dependence $\delta n \propto \exp(i \omega t)$, Eq.(7) can be reduced to
a  Hermitian equation [27]
$$
\omega^2 [\frac{d^2(n_0 \epsilon(n_0))}{d n_0^2}]^{-1} |f>=L|f>
%\frac{1}{m} \nabla(n_0\nabla f),
\eqno{(8)}
$$
where $|f>=\frac{d^2(n_0 \epsilon(n_0))}{d n_0^2}|\delta n>$,
$L=-\frac{1}{m} \nabla n_0\nabla$
 and the
 equilibrium density, $n_0$, is given by equation
$$
\mu=V_{ext}+\frac{d(n_0 \epsilon(n_0))}{d n_0},
\eqno{(9)}
$$
where $\mu$ is the chemical potential, in the region where $n_0(\vec{r})$ is
 positive and $n_0(\vec{r})=0$ outside this region.

We note here that $d^2(n_0 \epsilon(n_0))/d n_0^2$ is positive, since the sound velocity for the homogeneous case is given by $c^2=(n_0/m) 
d^2(n_0 \epsilon(n_0))/d n_0^2$.
\vspace{8pt}

\noindent
{\bf 3. Equation of state}
\vspace{8pt}

For the negative S-wave scattering length between the two fermionic species,
 $a<0$,
in the low-density regime, $k_F\mid a \mid \ll 1$, the ground state energy per
particle , $\epsilon(n)$, is well represented by an expansion in power of
$k_F \mid a \mid$ [32]
$$
\epsilon(n)=2 E_F[\frac{3}{10}-
\frac{1}{3 \pi} k_F \mid a \mid+0.055661 (k_F\mid a \mid)^2
-0.00914 (k_F\mid a \mid)^3+...],
\eqno{(10)}
$$
where $E_F=\hbar^2 k_F^2/(2 m)$ and $k_F=(3 \pi^2 n)^{1/3}$.
In the opposite regime, $a\rightarrow - \infty$
(the Bertsch many-body problem, quoted in
Refs.[33]), $\epsilon(n)$ is proportional to that of the non-interacting Fermi
 gas
$$
\epsilon(n)=(1+\beta)\frac{3}{10} \frac{\hbar^2 k_F^2}{m},
\eqno{(11)}
$$
where a  universal parameter  $\beta$ [10]  is estimated to be $\beta=-0.56$
 [34]. The universal limit [10,34-37] is valid at least in the case where the
width of the Feshbach resonance is large compared to the Fermi energy as in
 the cases of $^6Li$ and $^{40}K$.

 In the $a\rightarrow +0$ limit the system reduces to the
dilute Bose gas of dimers 
$$\epsilon(n)=E_F(-1/(k_F a)^2+a_m k_F/(6 \pi)+...,
)
\eqno{(12)}
$$
where $a_m$ is the boson-boson scattering length, $a_m \approx 0.6 a$ [38].

A simple interpolation of the form $\epsilon(n)\approx
E_F P(
k_F  a)$ with a smooth function $P(x)$ was considered in several papers.
In Ref.[21]  a [2/2] Pad\'{e} approximant
has  been proposed for the function
$P(x)$ for the  the case of negative $a$
$$
P(x)=\frac{3}{5}-2\frac{\delta_1\mid x \mid+\delta_2 x^2
}{1+\delta_3\mid x\mid+\delta_4 x^2},
\eqno{(13)}
$$
where $\delta_1=0.106103$, $\delta_2=0.187515$, $\delta_3=2.29188$,
$\delta_4=1.11616$.
Eq.(13) is constructed to reproduce the first four terms of the expansion (10)
 in
 the low-density regime and
  also  to exactly reproduce
 results of
 the recent Monte Carlo calculations [34], $\beta=-0.56$, in the  unitary limit,
$k_F a \rightarrow -\infty$.

 For the positive $a$ case ( the interaction is strong enough to form bound
 molecules
 with energy $E_{mol}$)  we have considered in Ref.[22]  a [2/2] Pad\'{e}
 approximant
$$
P(x)=\frac{E_{mol}}{2 E_F}+\frac{\alpha_1 x+\alpha_2 x^2}{1+\alpha_3 x+\alpha_4
x^2},
\eqno{(14)}
$$
where
 parameters $\alpha_i$ are fixed by two continuity conditions at large $x$,
$1/x\rightarrow 0$, and by two continuity conditions at small $x$,
 $\alpha_1=0.0316621$, $\alpha_2=0.0111816$, $\alpha_3=0.200149$,
and $\alpha_4=0.0423545$.

In Ref.[39] a Pad\'{e} approximation  has been considered   for the
 chemical potential. Authors of Ref.[28] have used a model for $P(x)$,
 interpolating the Monte Carlo results of Ref.[36] across the unitary limit
 and limiting behaviors for small $|x|$. We note here also the BCS mean-field
calculations of Ref.[28]. 
\vspace{8pt}

\noindent
{\bf 4. Schwinger variational principle}
\vspace{8pt}

The Schwinger variational principle  (SVP) [40-43] can be generalized to the 
case of
 Eq.(8).

Since the entire treatment is based on the equivalence of the SVP and the method of separable representation, we briefly describe this method. Let us consider the symbolic identity
$$
L=L L^{-1} L=\sum_{i,j} L|i><i|L^{-1}|j><j|L,
\eqno{(15)}
$$
where $|i>$ is a complete set. Truncating  the summation over the complete set we obtain a separable approximation
$$
L^{(q)}=\sum_{i,j}^q L|\chi_i>
d_{ij}^{-1} <\chi_j|L,
\eqno{(16)}
$$
where $d_{ij}=<\chi_i|L|\chi_j>$.  

We note that Eq.(16) represents an interpolation process, since
$L^{(q)}|\chi_i>=L|\chi_i>$ and $<\chi_j|L^{(q)}=<\chi_j|L$.

Substituting $L^{(q)}$ from Eq.(16) into Eq.(8), we obtain
$$
\omega^2 [\frac{d^2(n_0 \epsilon(n_0))}{d n_0^2}]^{-1} |f>=L^{(q)}|f>=
\sum_{i,j}^q L|\chi_i>
d_{ij}^{-1} <\chi_j|L|f>.
\eqno{(17)}
$$
We seek a solution of Eq.(17) in the form
$$
|f>=\sum_i^q c_i \frac{d^2(n_0 \epsilon(n_0))}{d n_0^2} L|\chi_i>,
\eqno{(18)}
$$
then $c_i$'s are defined from equations
$$
\sum_{k=1}^q B_{ik}(\omega^2) c_k=0,
\eqno{(19)}
$$
where
$$
B_{ik}(\omega^2)=<\chi_i|(\omega^2 L-L\frac{d^2(n_0 \epsilon(n_0))}{d n_0^2} L)
|\chi_k>
\eqno{(20)} 
$$
and frequencies $\omega$ are determined from the condition of vanishing of the determinant of the matrix $B_{ik}(\omega^2)$.
$$
det B_{ik}(\omega^2)=0                                                         
\eqno{(21)}
$$

 Defining the Schwinger functional $I_{SVP}[\chi] $ by
$$
I_{SVP}[\chi]= \frac{<\chi|L \frac{d^2(n_0 \epsilon(n_0))}{d n_0^2} L|\chi>}{<\chi| L|\chi>},
\eqno{(22)}
$$
 we get $I_{SVP}[f]=\omega^2$, where $f$ is the
 solution of Eq.(8). Introducing the function $\chi=\sum_{r=1}^q c_r \chi_r$,
 where $c_1,c_2, ... c_q
$ are $q$ variable parameters, we see that the functional $I_{SVP}$ is
 stationary if
$\partial I_{SVP}/\partial c_r=0$. The later equations coincide with Eqs.(19),
 and therefore the approximate solution of the Eq.(8), which is based on the SVP, is equivalent to the exact solution of the Eq.(8) with separable $L^{(q)}$.

Operator $L$ is clearly positive, which means that $<u|L|u>\geq0$ for all $u$,
 but is not positive definite, since $<u|L|u>=0$ for some $u\neq 0$.
It can easily be seen that
$$<u|(L-L^{(q)})|u>\geq 0,
\eqno{(23)}
$$
for all $u$.
Indeed,
$$
J=<(u+\sum_{i=1}^qc_i \chi_i)|L|(u+\sum_{j=1}^q c_j \chi_j)> \geq 0
$$
for all $c_i$.  We choose the $c_i$'s from the conditions
 $\partial J/\partial c_i=0$, and then obtain
$$
J=<u|(L-L^{(q)})|u>\geq0
$$
for all $q$.

Since the problem is solved by replacing $L$ by $L^{(q)}$ and since the
 operator $(L-L^{(q)})$ is positive, the SVP leads to the approximate lower
 bounds for $\omega$ up to the second order of $|L-L^{(q)}|$.

For the most interested case of $M=0$ modes, we can put in Eq.(16)
 $q=2$,$~$ 
$~$ $\chi_1=(x^2+y^2)$ and $\chi_2=z^2$, which give
$$
\omega^{SVP}_{\pm}=\frac{\omega_{\perp}}{\sqrt{2}} 
[\eta_{SVP} \pm \sqrt{\eta_{SVP}^2- 8 \lambda^2 (9 \zeta_{SVP}-1)}]^{1/2},
\eqno{(24)}
$$
where $\eta_{SVP}= 6 \zeta_{SVP}+ \lambda^2 (3 \zeta_{SVP}+1)$,
 $\zeta_{SVP}=-\int n_0^2 \tilde{x}^3 [\frac{\partial n_0}{\partial \tilde{x}}]^{-1} 
d \tilde{x}/\int n_0 \tilde{x}^4 d \tilde{x}$, 
 $\tilde{x}=\sqrt{x^2+y^2+\lambda^2 z^2}$ and  $\pm$ signs refer to the
 transverse and axial mode, respectively.

It is easy to show  that  Eq. (24) gives exact     
 solutions for frequencies of the breathing modes for the polytropic equation
 of state, $\epsilon(n)\approx n^{\gamma}$. 
In Ref.[24], on the basis of a generalization of the
 Hylleraas-Undheim method, we have constructed rigorous upper bounds to the
 collective
 frequencies for the radial and the axial breathing mode of the Fermi gas
 under harmonic confinement in the framework of the
 hydrodynamic theory
$$
\omega^{upper}_{\pm}=\frac{\omega_{\perp}}{\sqrt{5 \zeta_{upper}-9}}
[\eta_{upper} \pm \sqrt{
\eta_{upper}^2-8 \lambda^2 \zeta_{upper} (5 \zeta_{upper}-9)}]^{1/2},
\eqno{(25)}
$$
where $\eta_{upper}= (3+4 \lambda^2) \zeta_{upper} - (3+6 \lambda^2)$,
 $\zeta_{upper}=I_0 I_4/I_2^2,$ and
$I_l=\int \tilde{x}^l n_0(\tilde{x}) d\tilde{x}$.

We expect that the difference between $\omega^{SVP}_{\pm}$ and
$\omega^{upper}_{\pm}$ characterizes the error and is not very sensitive to 
the functional form of $\epsilon(n)$. From  Table 1 one can see
 that this difference is order of $10^{-5}$. We note that  the comparison with
 the
 scaling approximation, Eq.(6), shows that the absolute precision of the
 scaling approximation is about $10^{-3}$ that agrees with Ref.[31].

In Fig.1, we have  compared the hydrodynamic predictions for $\omega_+$
with experimental
data [17]. There is a very good agreement with experimental data [17] near
the unitary limit. We note here that two
 experimental results [17] and [16] (not shown in Fig.1)
 for $\omega_+$ are still
about  10\%  in disagreement  with
 each other, which is not fully understood yet.

To calculate $\zeta$'s, we have used the very fast converged  expansion of
Ref.[24]
$$
n_0(\vec{r})\approx (1-\beta V_{ext}(\vec{r}))^{1/(2-p)}
\sum_{i=0}^{l-1}\tilde{c}_{i} [V_{ext}(\vec{r})]^i, 
$$
where parameters $\beta$, $p$ and $\tilde{c}_i$ are fixed by requiring that
 $n_0(\vec{r})$ must satisfy a variational principle
 $\delta \int n_0 (V_{ext}+\epsilon(n_0))d^3 r=0$ with a subsidiary condition
 $\int n_0 d^3 r=N$.
\vspace{8pt}

\noindent
{\bf 4. Summary}
\vspace{8pt}

We have generalized the Schwinger variational method for the trapped strongly interacting atoms in hydrodynamic regime and we have constructed semi-analytical
 and extremely accurate formulae for hydrodynamic collective frequencies.
These formulae are very useful since they provide an easy and simple  quantitive tool for the analysis of experimental data for trapped condensed gases without relying on complex and extensive computations.

\pagebreak
 
Table 1. The transverse and axial frequencies
 in units of $\omega_{\perp}$ and $\omega_z$, respectively,
 in the BCS region as a function
 of the dimensional parameter
 $X=(N^{1/6} a/a_{ho})^{-1}$. The trap parameter $\lambda$ is assumed to be 
0.045613.
  The [2/2] Pad\'{e}
approximation
 of Refs.[21,22] is used for the energy per particle $\epsilon(n)$.\\

\begin{tabular}{lllllll}
\hline\hline
X
&$\omega^{upper}_+$
&$\omega^{SVP}_+$
&$\omega^{upper}_-$
&$\omega^{SVP}_-$ 
&$\omega^{s}_+$
&$\omega^{s}_-$ \\ \hline
-0.1
&1.8160
&1.8160
&1.5470
&1.5470
&1.8193
&1.5477 \\ \hline
-0.3
&1.8015
&1.8015
&1.5438
&1.5438
&1.8082
&1.5453 \\ \hline
-0.5
&1.7931
&1.7931
&1.5419
&1.5419
&1.8002
&1.5435 \\ \hline
-0.7
&1.7886
&1.7886
&1.5409
&1.5409
&1.7947
&1.5423 \\ \hline
-0.9
&1.7867
&1.7867
&1.5405
&1.5405
&1.7910
&1.5414 \\ \hline
-1.1
&1.7861
&1.7861
&1.5403
&1.5403
&1.7886
&1.5409 \\ \hline
-1.3
&1.7865
&1.7865
&1.5404
&1.5404
&1.7871
&1.5406 \\ \hline
-1.5
&1.7873
&1.7873
&1.5406
&1.5406
&1.7863
&1.5404 \\ \hline
-1.7
&1.7884
&1.7884
&1.5409
&1.5409
&1.7860
&1.5403 \\ \hline
-2.0
&1.7902
&1.7902
&1.5413
&1.5413
&1.7861
&1.5403 \\ \hline \hline
\end{tabular}

\pagebreak
\begin{figure}[ht]
\includegraphics{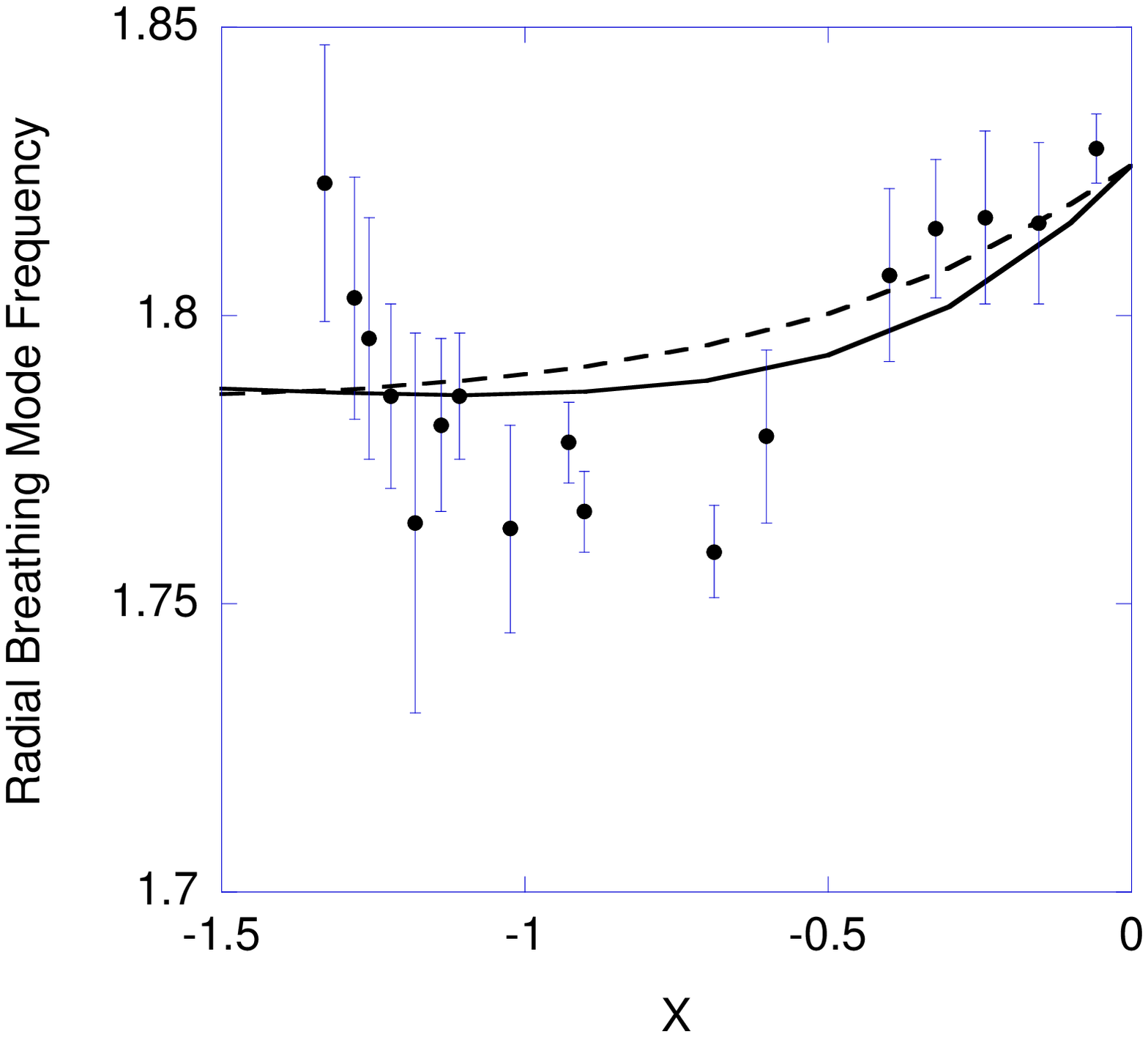}
\end{figure}
Fig. 1.
 Radial breathing mode frequency $\omega_+$  in the BCS region as a
function
 of the dimensional parameter $X=(N^{1/6} a/a_{ho})^{-1}$ (the solid line).
 The   dashed line  represent the scaling approximation, Eq.(6).
The  circular dots
 with error
 bars are the experimental results given by the Duke University group [17].
Everything is measured in units of $\omega_{\perp}$.
 
\pagebreak

{\bf References}
\vspace{8pt}

\noindent
[1]  Gehm M E  and  Thomas J E 2004
{\it Am. Sci.} {\bf 92} 238 

\noindent   
[2]  Regal C A,  Greiner M and  Jin D S  2004 {\it Phys. Rev. Lett.}
 {\bf 92} 040403

\noindent
[3]  Greiner M,  Regal C A  and  Jin D S 2003 {\it Nature} {\bf 426} 537 

\noindent
[4]  Stecker K E,  Patridge G B  and  Hulet R G 2003  
{\it Phys. Rev. Lett.} {\bf 91}
 080406 

\noindent
[5] Cubizolles J,  Bourdel T,  Kokkelmans S J J M F 
 Shlyapnikov G V  and
  Salomon C 2003 {\it  Phys. Rev. Lett.} {\bf 91} 240401 

\noindent
[6] Jochim S,  Bartenstein M,  Altmeyer A,  Hendl G,  Chin C,
  Denschlag J H ,
and
 Grimm R 2003 {\it Phys. Rev. Lett.} {\bf 91} 240402 

\noindent
[7] Jochim S,  Bartenstein M,  Altmeyer A,  Hendl G,  Riedl S,
 Chin C,
 Denschlag J H  and   Grimm R  2003 {\it Science} {\bf 302} 2101 

\noindent
[8]  Zwierlein M W,  Stan C A,  Schunck C H,  
Raupach S M F,  Gupta S,
 Hadzibabic Z,  Ketterle W 2003 {\it Phys. Rev. Lett.} {\bf 91} 250401 

\noindent
[9] Regal C A  and  Jin D S 2003 {\it Phys. Rev. Lett.} {\bf90}
 230404 

\noindent
[10]  O'Hara K M,  Hemmer S L,  Gehm M E,  Granade S R , and 
  Thomas J E 2002
{\it Science} {\bf 298} 2179

\noindent
[11]  Kinast J,  Turlapov A,  Thomas J E,  Chen Q,  Stajic J, and
  Levin K 2005 {\it Science} {\bf 307} 1296

\noindent
[12] Kinast J,  Turlapov A and Thomas J E 2005 {\it Preprint}
 cond-mat/0502507

2005 {\it Phys. Rev. Lett.} {\bf94} 170404

\noindent
[13] Bartenstein M,  Altmeyer A,  Riedl S,  Jochim S,  Chin C, 
 Denschlag J H and Grimm R 2004 {\it Phys. Rev. Lett.} {\bf92} 120401 

\noindent
[14] Kinast J,  Hemmer S L,    Gehm M E,  Turlapov A  and Thomas J E 2004
{\it Phys. Rev. Lett.} {\bf 92} 150402 

\noindent
[15] Bourdel T,  Khaykovich L,  Cubizolles J,  Zhang J,  Chevy F,
  Teichmann M,  Tarruell L,  Kokkelmans S J J M F and  Salomon C 2004
{\it Phys. Rev. Lett.} {\bf 93} 050401 

\noindent
[16] Bartenstein M,  Altmeyer A,  Riedl S,  Jochim S,  Chin C,
  Denschlag J H, and  Grimm R 2004 {\it  Phys. Rev. Lett.}
 {\bf 92} 203201 

2005 {\it Preprint} 2005 {\it Preprint} cond-mat/0412712

\noindent
[17] Kinast J,  Turlapov A, and  Thomas J E 2004 {\it Phys. Rev.}
 A {\bf 70}
 051401(R)

\noindent
[18] C. Chin C,  Bartenstein M,  Altmeyer A,  Riedl S,  Jochim S,
 Denschlag J H and   Grimm R 2004 {\it  Science} {\bf 305} 1128

\noindent
[19]
 Zwierlein M W,  Stan C A ,  Schunck C H,  Raupach S M F,  Kerman A J,
 and  Ketterle W 2004 {\ Phys. Rev. Lett.}  {\bf 92} 120403 

\noindent
[20]  Kinast J,  Turlapov A, and  Thomas J E, 2005 {\it Preprint} 
cond-mat/0503620

\noindent
[21] Kim Y E  and  Zubarev A L  2004 {\it Phys. Lett.}  A{\bf 327} 397 

\noindent
[22]  Kim Y E and  Zubarev A L 2004
{\it  Phys. Rev.} A
 {\bf 70} 033612 

\noindent
[23]  Kim Y E  and  Zubarev A L 2004 {\it Preprint}  cond-mat/0408283

\noindent
[24]  Kim Y E and  Zubarev A L 2005  {\it Preprint}  cond-mat/0502651.

\noindent
[25] Heiselberg H, 2004 {\it Phys. Rev. Lett.} {\bf 93} 040402 

\noindent
[26]  Menotti C, Pedri P and  Stringari S 2002 {\it Phys. Rev. Lett.}
 {\bf 89} 250402

\noindent
[27] Bulgak A and  Bertsch G F 2005  {\it Phys. Rev. Lett.}
 {\bf 94} 070401 

\noindent
[28]  Hu H,  Minguzzi A,  Liu Xia-Ji and  Tosi M P 2004
{\it Phys. Rev. Lett.}  {\bf 93} 190403 

\noindent
[29]  Manini N and  Salasnich L, 2005 {\it Phys. Rev.} A
 {\bf 71}  033625 

\noindent
[30]   Combescot R and  Leyronas X, 2004 {\it  Phys. Rev. Lett.}
 {\bf 93} 138901  

2004 {\it Europhys. Lett.} {\bf 68} 762 

\noindent
[31] Astrakharchik G E,  Combescot R,  Leyronas X and  Stringari S
2005 {\it Preprint} 
cond-mat/0503618

\noindent
[32]   Lenz W 1929 {\it Z. Phys.} {\bf 56} 778 

 Huang K and  Yang C N 1957
{\it Phys. Rev.}  {\bf105} 767 

 Lee T D  and  Yang C N  1957  {\it ibid} {\bf 105} 1119 

 Efimov V N  and  Amus'ya M Ya 1965 
{\it Sov. Phys. JETP} {\bf 20}, 388

\noindent
[33]  Baker  Jr. G A , 2001 {\it Int. J. Mod. Phys.} B{\bf15} 1314 

1999 {\it  Phys. Rev. C} {\bf
60}
054311 

\noindent
[34] Carlson J,  Chang S Y, Pandharipande V R and
  Schmidt K E 2003
{\it Phys. Rev. Lett.}  {\bf
 91}
050401 

\noindent
[35] Combescot R 2003 {\it Phys. Rev. Lett.}  {\bf 91} 120401 

\noindent
[36] Astrakharchik G E,  Boronat J,  Casulleras J  and
  Giorgini S 2004
{\it Phys. Rev. Lett.}  {\bf 93}, 200404 

\noindent
[37]  Ho T L  and  Mueller E J 2004 {\it  Phys. Rev. Lett.}
 {\bf 92} 160404 

 Ho T L 2004 {\it ibid} {\bf 92} 090402 

\noindent
[38] Petrov D S,  Salomon C, and  Shlyapnikov G V, 2004
{\it Phys. Rev. Lett.} {\bf 93} 090404 

\noindent
[39] Combescot R  and  Leyronas X, 2004 {\it Preprint}  cond-mat/0407388

\noindent
[40] Zubarev A L 1981  {\it The Schwinger Variational Principle in Quantum
 Mechanics} (Moscow: Energoatomizdat) [in Russian]

\noindent
[41] Zubarev A L 1976 {\it Sov. J. Part. Nucl.} {\bf 7} 215  

1978 {\it ibid} {\bf 9}, 188 

\noindent
[42]  Belyaev V B,  Podkopayev A P,  Wrzecionko J  and 
 Zubarev A L 1979
{\it J. Phys.}  B {\bf 12}, 1225  

\noindent
[43] Watson D K 1988 {\it  Adv. in Atom. Mol. Phys.}  {\bf 25} 221 

\end{document}